# Frequency comb generation at THz frequencies by coherent phonon excitation in Si

Muneaki Hase[1,2]★, Masayuki Katsuragawa[3], Anca Monia Constantinescu[1], and Hrvoje Petek[1]★

[1]Department of Physics and Astronomy, University of Pittsburgh, 3941 O'Hara Street, Pittsburgh, PA 15260, USA.
[2]Institute of Applied Physics, University of Tsukuba, 1-1-1 Tennodai, Tsukuba 305-8573, Japan.
[3]Department of Engineering Science, University of Electro-Communications, 1-5-1 Chofugaoka, Chofu, 182-8585, Tokyo, Japan.

**High-order nonlinear light-matter interactions in gases enable generation of x-ray and attosecond light pulses, metrology, and spectroscopy[1]. Optical nonlinearities in solid-state materials are particularly interesting for combining optical and electronic functions for high-bandwidth information processing[2]. Third-order nonlinear optical processes in silicon have been used to process optical signals with greater than 1 GHz bandwidths[2]. Fundamental physical processes for a Si-based optical modulator in the THz bandwidth range, however, have not yet been explored. Here we demonstrate ultrafast phononic modulation of the optical index of Si by irradiation with intense few-cycle femtosecond pulses. The anisotropic reflectivity modulation by the resonant Raman susceptibility at the fundamental frequency of the longitudinal optical (LO) phonon of Si (15.6 THz) generates a frequency comb up to 7th-order. All optical >100 THz frequency comb generation is realized by harnessing the coherent atomic motion of the Si crystalline lattice at its highest mechanical frequency.**

★Correspondence and requests for materials should be addressed to M. H. (mhase@bk.tsukuba.ac.jp) or H. P. (petek@pitt.edu).

The coherent modulation of electronic and vibrational nonlinearities in atoms and molecular gases by intense few-cycle pulses has been used to generate high-harmonic optical pulses in the soft X-ray and attosecond regime, as well as Raman frequency combs that span multiple octaves from the Terahertz to Petahertz (infrared to the vacuum ultraviolet) frequency regions[1,3-6]. In principle, similar high-order nonlinear processes can be excited efficiently in solids and liquids on account of their high nonlinear polarizability densities. In practice, however, the optical absorption, phase matching, and competition from other nonlinear processes such as white-light generation, limit applications of solid-state materials to low-order electronic processes, such as second-harmonic generation and optical rectification[7]. The nonlinear optical responses of solid surfaces, however, have been extended to high-harmonic generation[8] and multi-photon photoemission[9]. One might therefore anticipate that coherent vibrational Raman processes[10-12] in a highly nonlinear regime could also modulate light at multiple phonon frequencies in the 1-100 THz range.

The impulsive excitation of a crystalline lattice into a quantum optical coherent state[13] corresponding to oscillation at its highest mechanical frequency, the zone-centre LO phonon (coherent phonon), has been used to study nonlinear light-matter interactions, as well as electron-phonon and phonon-phonon coupling[14]. The coherent LO phonon spectroscopy and dynamics have been studied for insulators (diamond[15]), semiconductors (GaAs[16,17] and Si[18,19]), semimetals (Bi[20], Sb[20], and graphite[21]), metals[22], ferroelectrics[23], and organic crystals[24]. In many materials, including Si, the sudden generation of dense electron-hole (e-h) plasma both drives the coherent lattice vibration and induces lattice softening, potentially causing a structural phase transition[25]. The novel environment of such non-equilibrium plasmas could also promote highly nonlinear light-matter interactions absent in transparent materials.

Here, we explore the coherent phonon induced refractive index modulation of a Si(001) surface upon excitation at ≈ 397 nm (3.12 eV) in near-resonance[10] with the direct band gap of Si (≈3.4 eV) (Fig. 1)[19]. Through the anisotropic e-h pair generation and coherent Raman scattering, <1 nJ energy and ~10-fs duration laser pulses exert a sudden electrostrictive force on Si lattice launching coherent LO phonon oscillations at 15.6 THz frequency. With more than one order-of-magnitude larger amplitude[19] than in the case of non-resonant impulsive stimulated Raman excitation[18,24], the LO phonon oscillation strongly modulates the direct band gap of Si through the optical deformation potential[26,27]. The concomitant oscillatory change in the optical constants modulates the reflected probe light at the fundamental LO phonon frequency, generating a broad comb of frequencies at exact integer multiples of the fundamental frequency extending to beyond 100 THz. On the basis of an analytical model, we show that the simultaneous amplitude and phase modulation of the reflected light by the coherent lattice polarization at 15.6 THz generates the frequency comb.

Figure 2 shows the transient electro-optic (EO) reflectivity of the Si(001) sample upon excitation with a pump pulse of 1 mJ/cm² fluence, which is measured by scanning the pump-probe delay and recording the intensity difference between the orthogonal polarization components of the reflected light. Following an initial aperiodic electronic response, which follows the onset of the electrostrictive force[19], the signal oscillates with a period of ≈64 fs of the zone–centre coherent LO phonon and decays within several ps[18,19]. The coherent LO phonon response in Fig. 2 can be approximately simulated as a damped oscillator with an amplitude $A$, frequency $\omega_{LO}$ and its chirp $\eta$, delay-dependent relaxation time ($\tau_{LO} + \nu t$), and initial phase $\phi$,

$$\frac{\Delta R_{EO}(t)}{R_0} = A\exp[-t/(\tau_{LO} + \nu t)]\cos[(\omega_{LO} + \eta t)t + \phi]. \tag{1}$$

The LO phonon frequency, relaxation time, and phase depend on the photoexcited carrier density, which varies with both the pump fluence and the delay. The frequency shift and relaxation time change are manifestations of the complex self-energy for LO phonon interaction with photoexcited carriers (Table 1). The self-energy depends on the delay time as the carriers in the probing region relax to a thermal distribution and decay through transport[28]. The new finding from our experiment is evident in the residual of the fit in the Fig. 2 inset: in addition to dynamics represented by equation (1), the EO response oscillates at the 2nd and higher-order harmonics of the LO frequency with periods of 32, 21, and 16 fs, etc., down to 9 fs.

To further characterize the phonon-induced light modulation, in Fig. 3a we display the power spectrum of the transient reflectivity response obtained by Fourier transform (FT) of the time-domain signal in Fig. 2. The FT spectrum reveals that the modulated signal consists of an evenly spaced comb of frequencies dominated by the fundamental LO phonon oscillation at 15.6 THz and followed by a progression of its exact harmonics at 31.2, 46.8, 62.4, 78.0, 93.6 and 109.2 THz. Additional characteristics of the FT spectrum are: (i) the intensities alternate with even-orders being more intense than the odd; (ii) the linewidths increase with the order number $\beta$ approximately as $\Delta\Gamma \propto \beta$ (inset of Fig. 3a); and (iii) the lineshapes are asymmetric due to interference between the harmonic responses with the broad (>100 THz) pedestal of the fundamental response.

The evenly spaced FT line spectrum is reminiscent of a Raman frequency comb that can be generated through cascaded nonlinear interactions in transmission of *Frequency comb generation at THz frequencies by coherent phonon excitation in Si*intense laser pulses through transparent media[6]. In the case of the absorbing Si surface, however, the reflection depth and time are much less than an optical cycle, and therefore causally related multiple scattering through a cascading process cannot occur. Moreover, the frequency comb cannot be attributed to the phonon ladder climbing or the Raman overtone scattering, because these processes would produce anharmonic progressions. Therefore, we seek an explanation in the complex Raman polarization induced by the resonant pump pulse.

Our observation of the frequency comb is consistent with excitation of THz Raman polarization, which modulates the optical constants of Si through the optical deformation potential interaction. The electric field components $E_k$ and $E_l$ of the pump pulse in near-resonance with the direct band gap of Si exert a longitudinal force on the Si sample through a combination of charge density fluctuation and coherent Raman processes. The external field acts on the sample through the second-order nonlinear susceptibility $\chi^{(2)}_{jkl}(\omega_{LO};-\omega,\omega')$ to generate a rectified, longitudinal polarization, $P_j^{LO}(t) = \chi_{jkl}^{(2)} \cdot E_k(t)E_l(t)$, which is dominated by the coherent LO phonon response. Through the EO effect, $P_j^{LO}(t)$ modulates with the opposite phase the $E_k$ and $E_l$ components of the reflected probe light, generating collinear Stokes and anti-Stokes sidebands, $\omega_\mp$, via the nonlinear electro-optic Raman susceptibility, $\chi^{(2)}_{EO}(\omega_\mp;\omega,\mp\omega_{LO})$ (Fig. 1b).

To explain the observed frequency comb generation, we model the EO response by calculating the effect of $P_j^{LO}(t)$ on the incident field $E_0\cos\omega t$ of the probe. In response to $P_j^{LO}(t)$, the longitudinal component of the reflected probe electric field $E_{Pr}(t)$ has experienced a phase and amplitude modulation,

$$E_{Pr}(t) = \{a(t)\cos(\omega_{LO}t) + R_0\} E_0 \cos[\omega t - \delta(t)\cos(\omega_{LO}t + \varphi)], \qquad (2)$$

where $a(t)\cos(\omega_{LO}t) \propto \chi^{(2)}_{jkl} \cdot I$ describes the amplitude modulation (AM) and $\delta(t)\cos(\omega_{LO}t + \varphi) \propto \chi^{(2)}_{jkl} \cdot I$ describes the phase modulation (PM), with a pump intensity $I = |E_k E_l|$, $a(t) = a_0 e^{-t/\tau_{LO}}$, and $\delta(t) = \delta_0 e^{-t/\tau_{LO}}$. Other parameters are the static reflectivity, $R_0$, the optical carrier frequency, $\omega$, and the relative phase between the AM and PM, $\varphi$. The physical mechanism for the change in the real ($n$) and imaginary ($\kappa$) parts of the index of refraction is the band gap renormalization, i.e., the energy shift ($\Delta\varepsilon$) of the band edge by the optical deformation potential $\Xi$ (Fig. 1c), which is defined by $\Delta\varepsilon = \Xi|\Delta Q|/Q_0$[26], where $\Delta Q$ is the internal displacement (Si-Si) due to the optical phonon, and $Q_0$ is the static value of Si-Si bond length. The $|\Delta Q|$ dependence of the band gap and therefore the optical constants is expected for the $\Gamma_{25'}$ symmetry vibration of a crystal with an inversion symmetry.

The FT spectra of the simulated time-domain EO response of Si based on equation (2) are displayed in Fig. 3b. Together the AM and PM of the reflected probe light reproduce the frequency comb, consistent with the experiments. Specifically, the simulation reproduces: (i) the evenly spaced frequencies forming the comb at multiples of the exact LO phonon frequency; (ii) the line broadening as the order number increases; (iii) the even-odd order intensity alternation; and (iv) the alternating asymmetric lineshapes of the higher-orders. The asymmetry is best reproduced with the relative phase $\varphi \sim 0$ between AM and PM. Simulating the reflected field with PM only ($a_0$ = 0; Fig. 3b) generates even-orders only, whereas AM only ($\delta_0$ = 0) generates just the fundamental and 2-nd-order. Therefore, the simulations show that frequency comb results from the combined action of the amplitude and the phase modulation on the probe light.

An additional manifestation of the band gap renormalization is the dependence of the coherent phonon phase $\phi$ on the excitation density (see Table1). At low densities the coherent phonon phase $\phi$ is close to the impulsive limit ($\phi$ = 90°), indicating that the applied force is dominated by the Raman susceptibility[29]. By contrast, at high densities the band gap renormalization during the excitation brings the direct band gap of Si into resonance with the excitation light so that the anisotropic excitation of L-valley carriers exerts a displacive electrostrictive force ($\phi$ = 0°)[14,17,20]. We note that when replacing the AM and PM by sine functions in equation (2), as appropriate for the impulsive limit, the simulation reproduces the FT spectra obtained with 30 mW excitation corresponding to a lower photoexcited carrier density ($N \approx 0.5 \times 10^{20}$ cm$^{-3}$).

In summary, we have discovered a new approach for ultrafast light modulation by coherent phonon excitation. The THz polarization at the zone- centre LO phonon frequency (15.6 THz) of Si acts through the complex Raman susceptibility on the phase and amplitude of the reflected light. The resulting frequency comb that is impressed on the reflected light extends up to 7th-order (109.2 THz) of the highest mechanical frequency of the driving polarization, and is limited in bandwidth by the duration of the probe pulse. The characteristic spectral features of the frequency comb can be attributed to the simultaneous action of the amplitude and phase modulation on the electric field of the probe light during the process of reflection from the Si surface. The strong nonlinear interaction that creates the effective amplitude and phase modulation on a small length scale compared to the optical wavelength is caused by the renormalization of the optical band gap of Si by the resonant anisotropic e-h pair excitation. The frequency and shape of the comb generated by coherent LO phonon can be controlled by the pump fluence (magnitude of $a_0$ and $\delta_0$) and the choice of the material (the values of $\tau_{LO}$ and $\omega_{LO}$). Therefore, our approach is only limited by the bandwidths of the excitation and probe light.

## Methods

### Ultrafast spectroscopy

The anisotropic transient reflectivity of *n*-doped ($1.0\times10^{15}$ cm$^{-3}$) Si(001) wafer was measured in air at 295 K by the EO sampling technique[16,17]. Nearly collinear, pump and probe beams [397 nm (3.12 eV) centre wavelength] were overlapped at a $7.2\times10^{-7}$ cm$^2$ spot on the sample. The frequency doubled Ti:sapphire laser oscillator with an average power of 60 mW and a 70 MHz repetition rate generated $N \approx 1.0\times10^{20}$ cm$^{-3}$ carriers as estimated from the absorption coefficient $\alpha = 1.2\times10^5$ cm$^{-1}$ at 397 nm[30]. The carrier density is one to two orders-of-magnitude less than thresholds for optical damage and optically induced melting of Si[25]. The polarization of the pump beam was set to [110] direction, while that of the probe was [100][19]. The reflected probe beam was analyzed into polarization components parallel and perpendicular to that of the pump and each was detected with a photodiode. The resulting photocurrents were subtracted and after amplification their difference [$\Delta R_{eo}/R_0 = (\Delta R_k - \Delta R_l)/R_0 = (|E_k|^2 - |E_l|^2)/|E_0|^2$] was recorded versus the pump-probe delay. The delay was scanned over 10 ps and averaged for 20,000 scans by using an oscillating retroreflector with a 20 Hz scan frequency.

### Numerical simulation

The generation of the frequency comb is simulated by calculating the modulation of the probe light through the complex Raman susceptibility by coherent LO phonon oscillation (Fig. 1a,b). The THz Raman polarization modulates the complex index of refraction $\tilde{n} = n + i\kappa$. In Si at 3.12 eV the real and imaginary parts are $n = 5.57$ and $\kappa = 0.387$, therefore $n \gg \kappa$[30]. The derivatives of $n$ and $\kappa$ with respect to photon energy ($\hbar\omega$), however, are different, with $d\kappa/d\omega$ being larger than $dn/d\omega$ near the band gap[30]. Consequently, $\Delta n \gg \Delta\kappa$ does not hold, and both $\Delta n$ and $\Delta\kappa$ contribute to the light modulation. In equation (2) we define $\delta_0 = (2\pi L/\lambda_0)\cdot\Delta n = 1.30\Delta n$, where $\Delta n \propto \Delta\varepsilon \propto$ $|ΔQ|$[26, 27], $L ≈ 1/α = λ_0/4πκ = 82$ nm is the optical penetration depth, and $λ_0 = 397$ nm the laser wavelength. For the reflectivity, the static value $R_0 = 0.486$ was used[30]. Furthermore, from the Fresnel's coefficient with nearly surface normal incidence, $a_0 = 2Δκ/(n+1)^2 = 0.0463Δκ$ if $Δκ > Δn$, where $Δκ ∝ Δε ∝ |ΔQ|$. The ratio of magnitudes of the phase to the amplitude modulation used in the simulation thus becomes $δ_0/a_0 ≈ 28$. By the Jacobi-Anger identity, we can expand equation (2) as:

$$\cos[δ(t)\cos(ω_{LO}t + φ)]$$
$$= J_0(δ(t)) - 2J_2(δ(t))\cos(2ω_{LO}t + 2φ) + 2J_4(δ(t))\cos(4ω_{LO}t + 4φ) - \cdots,$$

$$\sin[δ(t)\cos(ω_{LO}t + φ)]$$
$$= 2J_1(δ(t))\cos(ω_{LO}t + φ) - 2J_3(δ(t))\cos(3ω_{LO}t + 3φ) + 2J_5(δ(t))\cos(5ω_{LO}t + 5φ) - \cdots,$$

where $J_i(z)$ is the Bessel function of the first kind of the order of $i$ (= 0, 1, 2, $\cdots$). Using these relations, equation (2) can be rewritten as,

$$E_{Pr}(t) = E_0[a(t)\cos(ω_{LO}t) + R_0] ×$$
$$\{J_0(δ(t))\cos(ωt) + J_1(δ(t))\sin[(ω+ω_{LO})t + φ] + J_1(δ(t))\sin[(ω-ω_{LO})t - φ]$$
$$- J_2(δ(t))\cos[(ω+2ω_{LO})t + 2φ] - J_2(δ(t))\cos[(ω-2ω_{LO})t - 2φ]$$
$$- J_3(δ(t))\sin[(ω+3ω_{LO})t + 3φ] - J_3(δ(t))\sin[(ω-3ω_{LO})t - 3φ]$$
$$+ J_4(δ(t))\cos[(ω+4ω_{LO})t + 4φ] + J_4(δ(t))\cos[(ω-4ω_{LO})t - 4φ]$$
$$+ \cdots \}. \qquad (3)$$

In our model, the pump pulse excites coherent LO phonons, creating the THz polarization at the phonon frequency $ω_{LO}$. This polarization modulates the amplitude and phase of the probe pulse via the complex Raman susceptibility according to equation (2). The oscillatory observable in the EO sampling measurements is the longitudinal phononic part of the difference between $|E_k(t)|^2$ and $|E_l(t)|^2$ components, which is expressed by $|E_{Pr}(t)|^2/E_0^2$;

$$|E_{Pr}(t)|^2/E_0^2 ≈ 2a(t)R_0\left(\frac{1}{2}J_0^2 + \frac{3}{2}J_1^2 + J_2^2 + J_3^2 + J_4^2 - J_0J_2 - J_1J_3 - J_2J_4\right)\cos(ω_{LO}t)$$
$$+ 2R_0^2\left(\frac{1}{2}J_1^2 - J_0J_2 - J_1J_3 - J_2J_4\right)\cos(2ω_{LO}t) + \frac{1}{4}a(t)^2 J_0^2\cos(2ω_{LO}t)$$

$$+2a(t)R_0\left(\frac{1}{2}J_1^2 + \frac{1}{2}J_2^2 - J_0J_2 - 2J_1J_3 + J_0J_4 - J_2J_4\right)\cos(3\omega_{LO}t)$$

$$+2R_0^2\left(\frac{1}{2}J_2^2 - J_1J_3 + J_0J_4\right)\cos(4\omega_{LO}t)$$

$$+2a(t)R_0\left(\frac{1}{2}J_2^2 + \frac{1}{2}J_3^2 - J_1J_3 + J_0J_4 - J_2J_4\right)\cos(5\omega_{LO}t)$$

$$+2R_0^2\left(\frac{1}{2}J_3^2 - J_2J_4\right)\cos(6\omega_{LO}t)$$

$$+2a(t)R_0\left(\frac{1}{2}J_3^2 + \frac{1}{2}J_4^2 - J_2J_4\right)\cos(7\omega_{LO}t) + \cdots. \qquad (4)$$

In equations (3) and (4), we neglect the terms higher than the 5th order ($J_5$, $J_6$, $\cdots$) because of their $10^{-2}$–$10^{-3}$ smaller values than the lower-orders. Including the 5th and 6th-order terms in equations (3) and (4), generates responses up to the 12th order of the LO phonon frequency ($\approx$ 200 THz), which is beyond the bandwidth of our laser. In equation (4) we note that the amplitude for the odd-order terms include the AM "$a(t)$", whereas the even-order (except the 2nd order) terms do not; this difference explains the odd–even order intensity alternation. Furthermore, for pure AM, i.e., when $\delta_0 = 0$ then $J_0 = 1$, and $J_i$ ($i$ = 1, 2, 3, $\cdots$) =0, and only the 1st and 2nd-orders appear (see Fig. 3b), because only these orders include the $J_0^2$ term in equation (4). Finally, we note that a simple argument with AM being expressed by $E_{Pr} = E_0 \Delta\kappa \propto |\Delta Q| \propto |\cos(\omega_{LO}t)|$, and PM by $E_{Pr}(t) = E_0 \cos(\Delta n) \propto E_0\cos(|\cos(\omega_{LO}t)|) \propto E_0\left[1-\frac{1}{2}\cos^2(\omega_{LO}t)+\cdots\right]$ gives the same qualitative behaviour as the calculated AM and PM contributions from the solution of the Bessel equations (FT spectra in Fig. 3b).

**Table 1. Parameters obtained by the fit of the transient EO response at various pump fluences.**

| Pump fluence (mJ/cm$^2$) | Frequency (THz) | Relaxation time (ps) | Chirp (ps$^{-2}$) | Initial phase (degree) |
|---|---|---|---|---|
| 0.08 | 15.613 | 3.62 | 0.00381 | 68.0 |
| 0.50 | 15.596 | 2.12 | 0.00629 | 60.7 |
| 1.00 | 15.587 | 1.31 | 0.01120 | 10.0 |

**Acknowledgements** The authors acknowledge M. Kitajima for stimulating discussions. This work was supported in part by NSF under grant CHE-0650756.

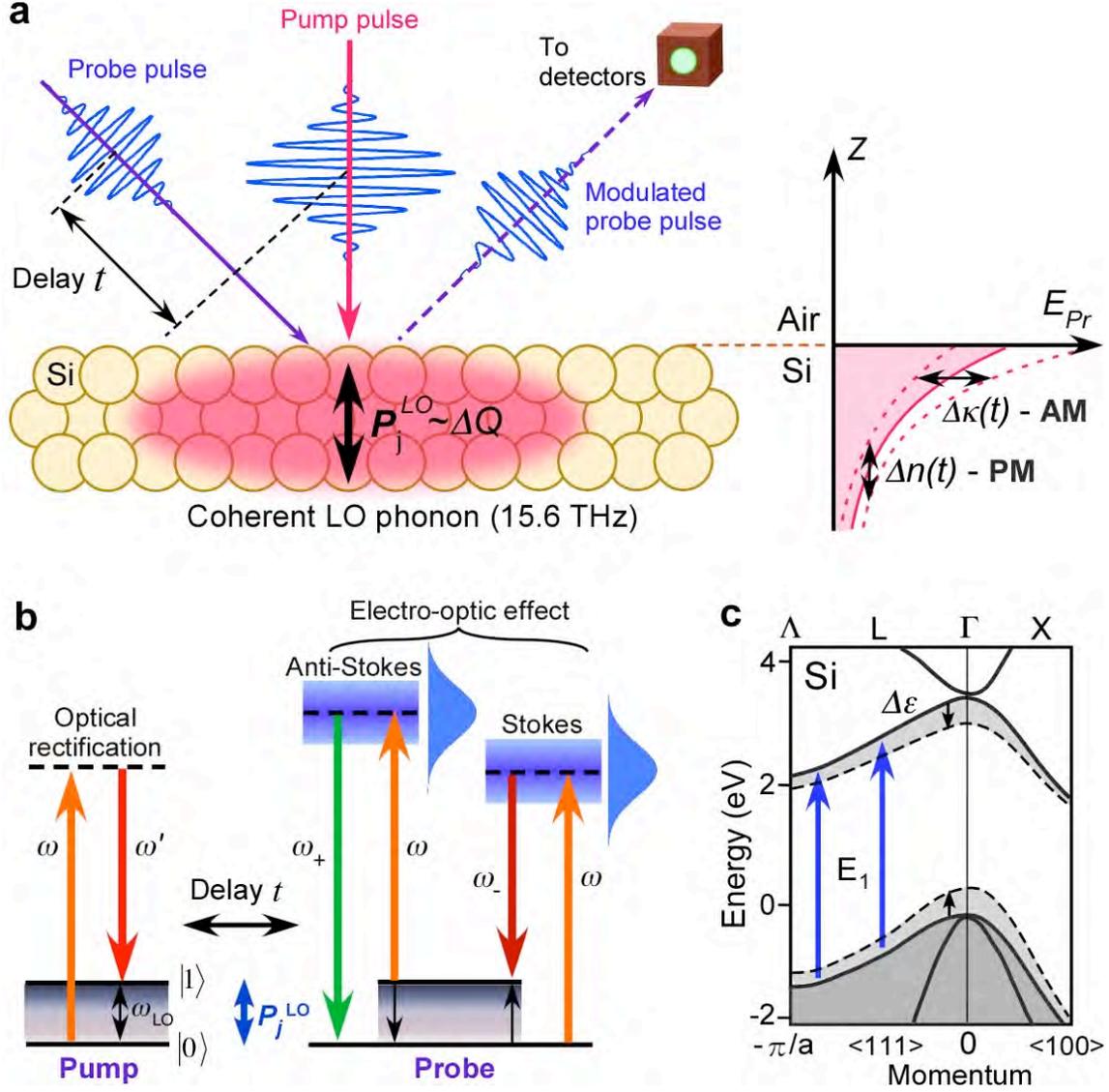

**Figure 1: Schematic of the refractive index modulation in Si. a**, Pump-probe configuration and modulation of the probe pulse by the coherent THz polarization ($P_j^{LO}$) after the photoexcitation; $t$ represents the time delay between the pump and probe pulses. The right panel describes penetration of the probe electric field, $E_{pr}$, which is modulated by the complex index changes of $\Delta n$ (:PM) and $\Delta \kappa$ (:AM) ($\propto P_j^{LO} \propto \Delta Q$). **b**, The sequence of excitation and probing actions can be described by processes of optical rectification $\chi_{jkl}^{(2)}(\omega_{LO};-\omega,\omega')$ and electro-optic effect $\chi_{EO}^{(2)}(\omega_{\mp};\omega,\mp\omega_{LO})$, during which the Stokes and anti-Stokes sidebands are generated within the bandwidth of the probe pulse (Gaussian shape with blue colour). **c**, Band gap renormalization ($\Delta\varepsilon$) in the energy bands of Si via the optical deformation potential. Photoabsorption at 3.12 eV dominates the $E_1$ critical point (3.396 eV) along the *L* valleys.

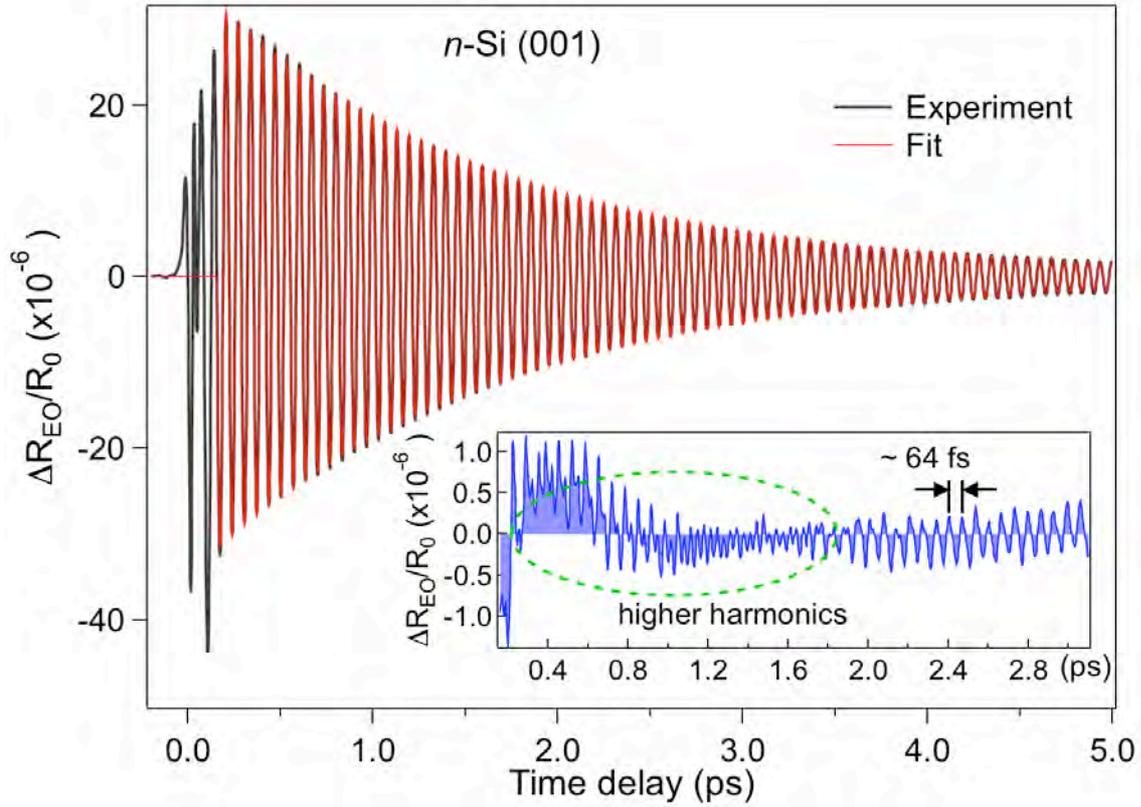

**Figure 2: Coherent phonon oscillations.** Transient EO reflectivity was observed with 397 nm (3.12 eV) excitation at 60 mW ($N \approx 1.0 \times 10^{20}$ cm$^{-3}$). The black line represents the experimental data and the red line is the fit to equation (1). The fit parameters are given in Table I, for several photoexcited carrier densities. The inset is the residual of the fit, showing

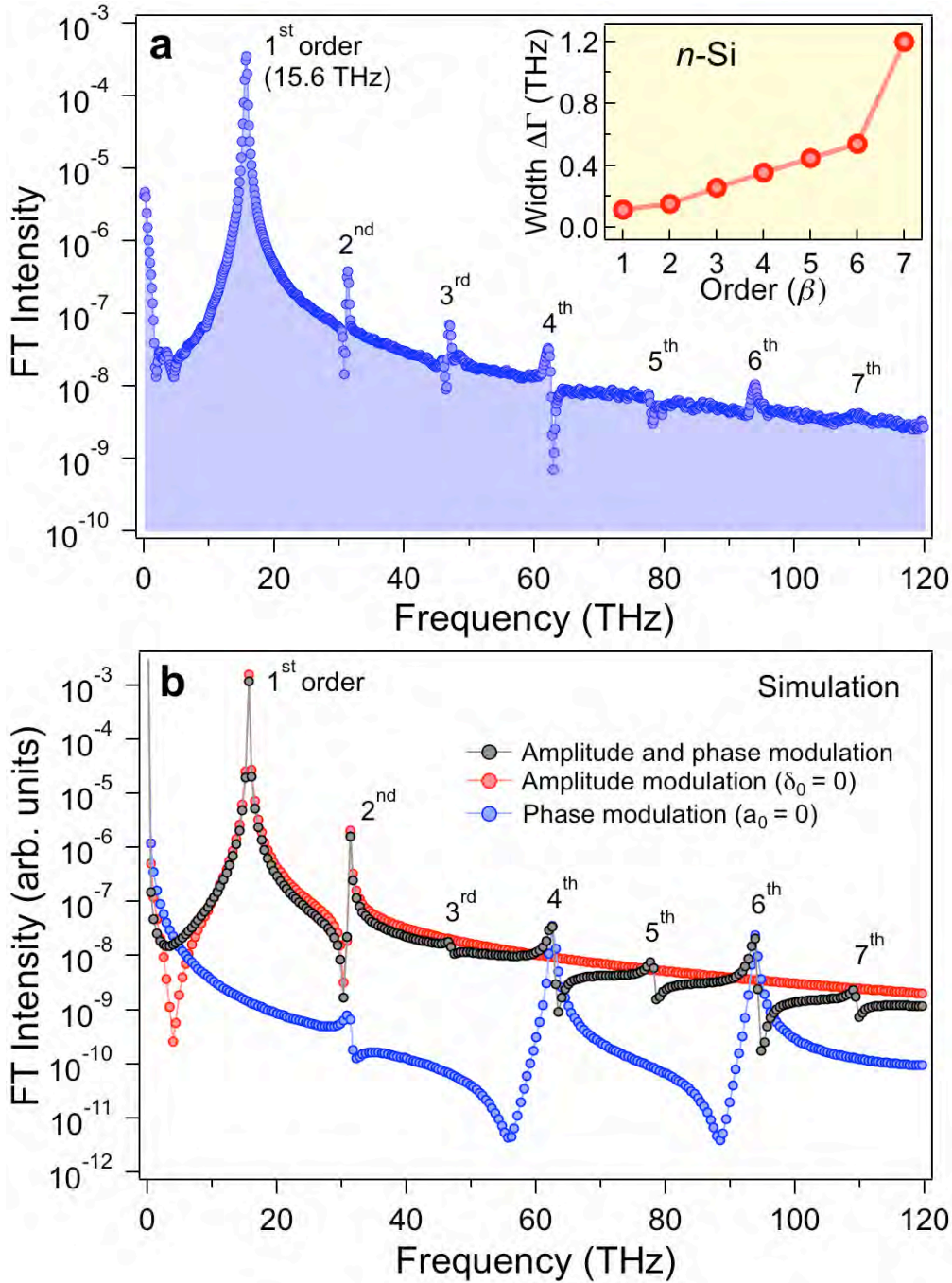

**Figure 3: FT spectrum obtained from the transient anisotropic reflectivity. a**, FT spectrum of the reflectivity signal of Fig. 2. The inset represents the width of the higher order terms as a function of their order $\beta$. **b**, FT spectra of the reflectivity signal calculated using the model in equations (2) – (4). The simulated spectrum with the amplitude and phase modulation with $\varphi \sim 0$ in equation (2) acting together or separately. The experimental frequency comb generation can only be reproduced by including both the phase and the amplitude modulation, as described in the main text.